# Etch-stop method for reliably fabricating sharp yet mechanically stable scanning tunneling microscope tips

Gobind Basnet, J. Kevin Schoelz, Peng Xu, Steven D. Barber, Matthew L. Ackerman, and Paul M. Thibado[a), b)]

Department of Physics, University of Arkansas, Fayetteville, Arkansas 72701

[a)]American Vacuum Society member.

[b)]Electronic mail: thibado@uark.edu

An extension of the direct-current, double-lamella drop-off technique for electrochemically etching tungsten scanning-tunneling-microscope tips is presented. The key fabrication step introduced here is the use of an etch stop as a simple but accurate way to optimize the contact area between the etchant and the wire. By restricting the etching process, the final cone angle of the tips can be made sharp and mechanically stable without a lot of finesse from the STM tip maker.

# I. INTRODUCTION

The fabrication of scanning-tunneling-microscope (STM) tips through electrochemical etching of tungsten (W) wire is a well-developed art.[1-3] Several implementations of this general scheme have been found to reliably produce a small enough tip apex radius to be capable of high-resolution imaging. However, in the pursuit of atomic sharpness, the importance of mechanical stability has sometimes been overlooked.[4] For example, an STM tip with an extremely long taper (i.e., a very small cone angle) will vibrate when driven by the piezoelectric scanner of the STM, making atomic-resolution imaging impossible. In fact, the most stable configuration for an STM tip is a pyramidal arrangement of atoms, with one atom at the apex, three atoms below that, six atoms below that, and so on, yielding an overall cone angle of about 60°. This type of tip has been imagined and is what has led to the use of single-crystal W(111) oriented wire.[6] Correspondingly, the overall stability of an STM tip can be primarily determined by its shape, as characterized by an optically measured aspect ratio or cone angle.[5] In the ideal situation, a sufficiently large cone angle should be obtained, while requiring a minimal amount of skill from the STM tip maker.

A particularly dependable electrochemical etching technique, known as the double-lamella method, regularly produces sharp STM tips but also permits control over a wide range of shapes.[7] In this setup, the electrolytic etching solution is suspended in two parallel metal loops held at a fixed potential difference, while the W wire is oriented vertically through the center of both loops, completing the etching circuit. When the wire is etch through at the upper loop, the bottom end of the wire drops off under the influence of gravity, the circuit is broken to the lower loop, and etching terminates almost immediately. In this setup, the final cone angle characteristics are largely determined by the thickness of the etching lamella (i.e., the contact area between the etchant and the wire), which are a function of the loop size and the surface energy of the solution. Thinner lamellae produce tips with large cone angles, but these lamellae are unfortunately not very robust. They tend to repeatedly break, resulting in longer etch times and requiring the technician to restore the lamella to the same location on the wire each time. Thicker lamellae are more durable and etch more quickly, but they create much smaller cone angles due to increased contact area of the etchant along the axis of the wire. Therefore, a method is sought which would reduce this contact area without reducing the thickness of the etching lamella, in order to reliably produce tips that are both sharp and mechanically stable.

In this study, we report the first use of an etch stop in the electrochemical etching of tungsten STM tips. An etch stop is a material that is inert with regards to the etchant, and this powerful technique has been widely employed in other fields to control the rate of etching along a surface.[8,9] We demonstrate that applying an etch stop to protect a section of the W wire can result in larger tip cone angles, and thus enhanced mechanical stability, while still using a thick, stable lamella.

## II. DESIGN AND FABRICATION

The STM tips in this study were etched using a double-lamella, gravity-switch system, shown schematically in Fig. 1(a). Starting at the top is an inverted stainless steel STM tip holder (Omicron). Prior to this orientation, a 5-mm length of 0.15-mm diameter tantalum (Ta) wire was placed into the tip holder followed by a 10-mm length of 0.25-mm diameter W wire. By initially using the normal STM tip holder, one eliminates extra handling steps after production. The Ta is relatively soft, so it deforms to securely hold the W wire when it is pressed into the tip holder. The Ta wire is then trimmed to a couple millimeters and bent back so as not to interfere with the STM operation. This procedure also permits one to reuse the tip holder by removing the wires later simply by grabbing and pulling on both wires simultaneously. After securing the W wire to the holder, a thin layer of etch stop about 0.05 mm thick is applied to the W wire by submerging it into the etch stop solution. This starts from the free end and continues up to about 3 mm away from the tip holder. Care must be taken to ensure that the coating is uniform around the entire circumference of the W wire-etch stop boundary to guarantee symmetric etching of the tip. To this end, it is helpful to completely dry the etch stop immediately after application using a heat gun. A small portion of the etch stop, at the free end of the W wire, is then mechanically uncoated so that current will be able to flow through the etching circuit. A photograph of the prepared wire with etch-stop material attached (blue), and the tip holder with the Ta wire is shown in Fig. 1(b).

Next the tip holder is attached to a magnet that keeps it in place during the etching process. The W wire extends downward through two gold (Au) loops located 5 mm apart and oriented so that their faces are parallel and lie in horizontal planes [see Fig. 1(a) again]. The top loop has a diameter of 15 mm, while the bottom loop has a diameter of only 7 mm. The position of the two rings is controlled using a micromanipulator (not shown) which offers coarse adjustment in the $x$, $y$, and $z$ directions, as well as a fine hydraulic control for the $z$ direction (tip holder is stationary). This allows the operator to maintain etching at precisely the desired location with minimal vibration. An etching solution (8 g of sodium hydroxide in 100 mL of deionized water) is suspended in both of the Au loops by briefly submerging them, and a direct-current power supply is attached to both rings to apply a DC bias of 8 V between them. Hydrogen bubbles are formed at the lower ring and no bubbles are formed at the upper ring where etching takes place. This allows unobstructed real-time observation of the etching process and produces more uniform etching. A 30x microscope (not shown) is used to monitor the position and thickness of the etching lamella on the W wire. At the completion of the process, the bottom end of the W wire drops off due to the influence of gravity, breaking the circuit and ending the etching.

## III. RESULTS AND DISCUSSION

More than 60 STM tips were fabricated for this study; at least 20 for each of the three different etch stop positions illustrated in the top row of Fig. 2. In the first position, the top of the etch stop was aligned with the top of the etching meniscus, resulting in a contact length of 0.0 mm between the wire and etchant as shown schematically in the top part of Fig. 2(a). In this configuration the average etching time was 90 min before the bottom wire dropped. Optical micrographs of representative STM tips at 10x magnification are shown beneath the schematic. Due to the excessive etching times required, the structures observed in this column are dominated by undercutting of the etch stop. The etchant penetrates behind the etch stop, resulting in a variety of messy, useless tip profiles.

In the second position, depicted schematically in Fig. 2(b), the top of the etch stop is 0.6 mm from the top of the etching meniscus. The typical etch time at this position was about 30 min. Below the schematic, three representative STM tips for this arrangement are shown in optical micrographs at 100x magnification. The tip shape was very reproducible, and the undercutting that was previously present is negligible. The tips are long and sharp, with a characteristic cone angle of 20° or less about 85% of the time. While this method is an improvement over the first, the cone angles are similar to what one achieves without using the etch stop. We have found that the majority of these tips are not mechanically stable enough for atomic-resolution STM imaging.

In the final position, shown schematically in Fig. 2(c), the top of the etch stop is placed half way between the above two described positions or about 0.3 mm below the top of the etching meniscus, and the typical etch time is about 40 minutes. Under the schematic, three representative tips are displayed at 100x magnification. Again, sharp tips were consistently produced, but 80% of these tips had cone angles greater than 20°. At these cone angles, we have found that they are mechanical stability enough for atomic-resolution STM. A typical high-quality atomic-resolution STM image of graphite acquired using the STM tips being made under this condition is shown in Fig. 2(d) (see figure caption for STM system and settings). Also, at less than half the etching time of the first configuration, undercutting of the etch stop is still negligible. Furthermore, because a thick etching lamella is used, the etchant rarely needs to be rewetted, saving the tip maker time and frustration.

## IV. CONCLUSION

Double-lamella techniques are extremely advantageous in the production of sharp STM tips. However, the method can also be time-consuming and greatly depend upon the skill of the technician, or else it will produce tips that are mechanically unstable. By combining the double-lamella technique with an etch stop methodology, we showed that it is possible to increase the likelihood of producing sharp yet mechanically stable STM tips, while simultaneously making it easier for the operator.

## ACKNOWLEDGMENTS

This work was supported in part by NSF under grant DMR-0855358 and ONR under grant N00014-10-1-0181.

**FIGURE CAPTIONS**

FIG. 1. (Color online) (a) Schematic of the experimental double-lamella electrochemical tip etching setup, including the application of an etch stop. (b) Photograph of the W wire in an inverted tip holder after it has been fully prepared for etching.

FIG. 2. (Color online) (a) A schematic illustrates how the etch stop was configured relative to the etchant for a series of trial tips. Representative optical micrographs of the results at 10x magnification are displayed below the diagram. (b,c) Show the same, but at 100x magnification for the optical micrographs. (d) Atomic-resolution STM image of graphite acquired using the tip made with the configuration in (c). [This image was acquired using an Omicron UHV LT-STM operated at room temperature, a tip setpoint bias of 0.1 V and a constant tunneling setpoint current of 0.1 nA.]

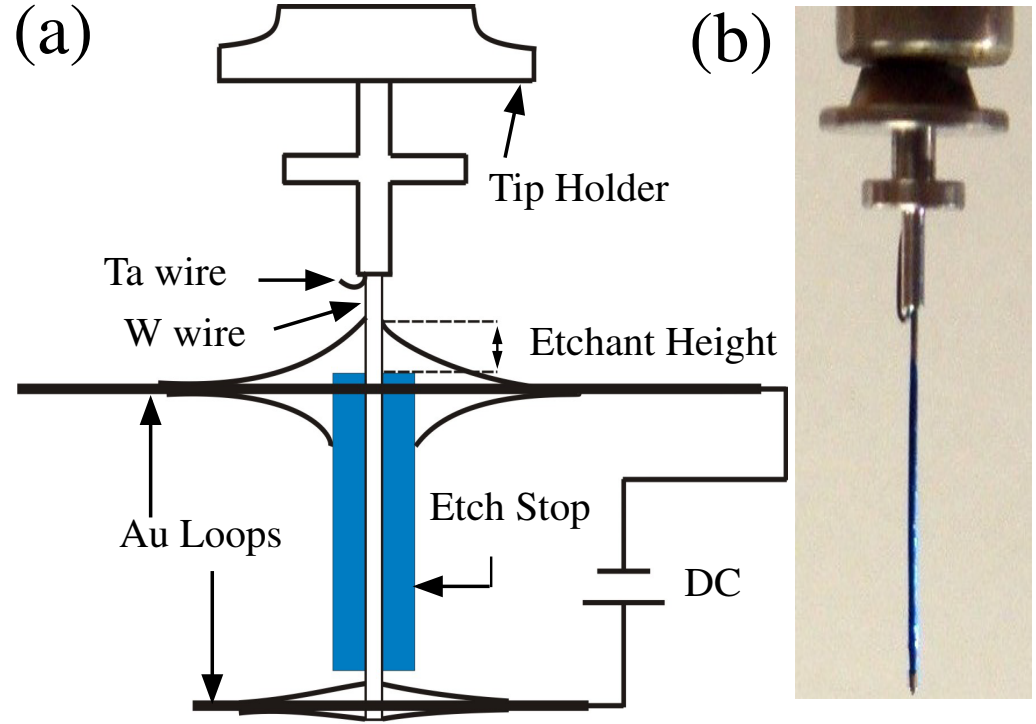

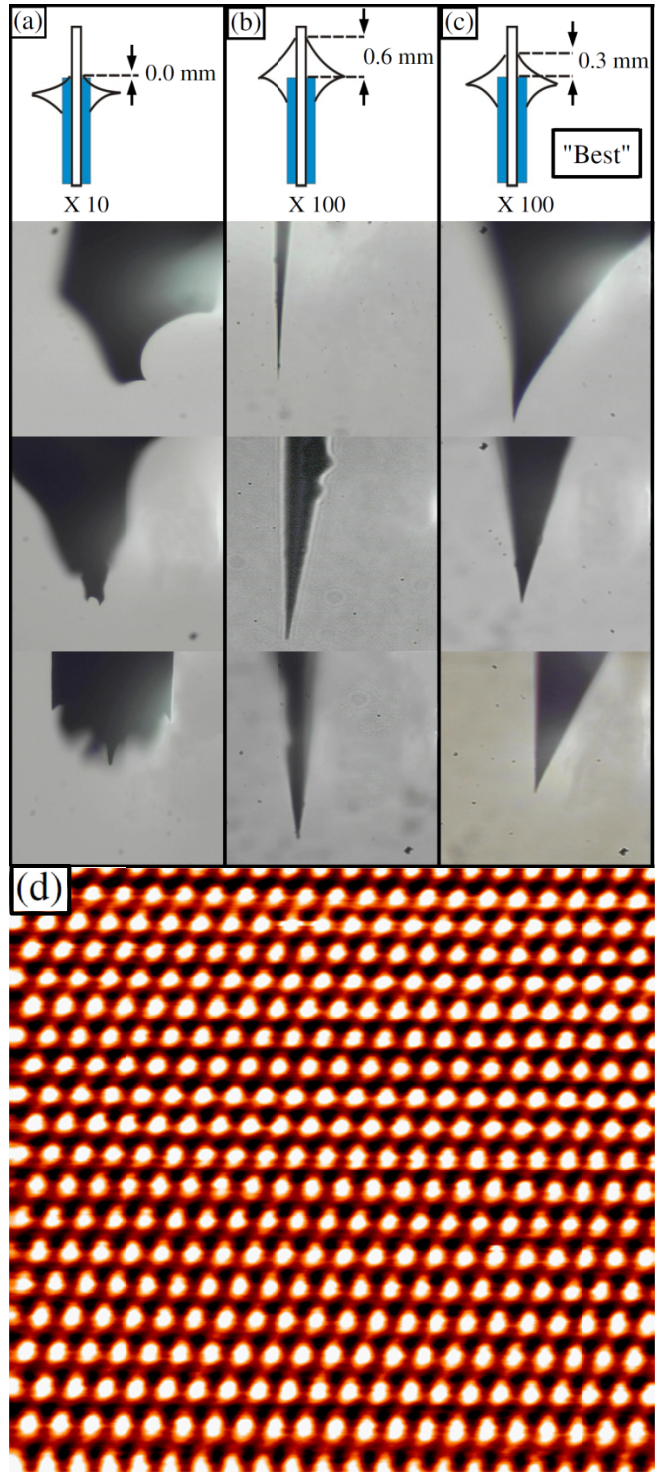